\documentclass[reprint,secnumarabic,amsmath,amssymb, nobibnotes, aps, prl,superscriptaddress]{revtex4-2}
\usepackage{graphicx}
\usepackage{xcolor}
\usepackage[nolist]{acronym}
\usepackage{microtype}
\usepackage{placeins,float}

\newcommand{\sx}{\sigma_{\text{x}}}
\newcommand{\Sx}{\Sigma_{\text{x}}}
\newcommand{\sy}{\sigma_{\text{y}}}
\newcommand{\sz}{\sigma_{\text{z}}}
\renewcommand{\sp}{\sigma_{+}}
\newcommand{\sm}{\sigma_{-}}
\newcommand{\Tr}{\operatorname{Tr}}
\newcommand{\ind}[1]{_{\text{#1}}}
\newcommand{\ii}{\text{i}}
\newcommand{\e}{\text{e}}
\renewcommand{\d}{\text{d}}
\renewcommand{\Re}{\operatorname{Re}}

\DeclareMathOperator*{\argmin}{arg\,min}
\DeclareMathOperator*{\cov}{cov}

\begin{document}

\title{Connection between memory performance and optical absorption in quantum reservoir computing}%

\author{Niclas Götting}%
\affiliation{Institute for Physics, Faculty V, Carl von Ossietzky University Oldenburg, 26129 Oldenburg, Germany}
\author{Steffen Wilksen}%
\affiliation{Institute for Physics, Faculty V, Carl von Ossietzky University Oldenburg, 26129 Oldenburg, Germany}
\author{Alexander Steinhoff}%
\affiliation{Institute for Physics, Faculty V, Carl von Ossietzky University Oldenburg, 26129 Oldenburg, Germany}
\author{Frederik Lohof}%
\affiliation{Institute for Theoretical Physics, University of Bremen, 28359 Bremen, Germany}
\author{Christopher Gies}%
\affiliation{Institute for Physics, Faculty V, Carl von Ossietzky University Oldenburg, 26129 Oldenburg, Germany}
\date{\today}%

\begin{abstract}
Quantum reservoir computing (QRC) offers a promising paradigm for harnessing quantum systems for machine learning tasks, especially in the era of noisy intermediate-scale quantum (NISQ) devices.
While information-theoretical benchmarks like short-term memory capacity (STMC) are widely used to evaluate QRC performance, they fail to provide insights into the physical mechanisms underlying these quantum neural networks.
We establish a quantitative connection between the optical absorption spectrum of a quantum reservoir and its memory performance, revealing that optimal STMC aligns directly with maximal absorption, providing a physical explanation for the previously reported ``sweet-spot" behavior in QRC performance as a function of dissipation.
This connection bridges quantum information theory with experimentally accessible physical properties, opening pathways for targeted engineering of quantum reservoir computers with optimized performance for specific tasks.
\end{abstract}

\maketitle

\section{Introduction}
In recent years, quantum systems emerged as an intriguing new platform for \ac{rc}, a special type of \ac{rnn} \cite{elman_finding_1990,jordan_serial_1986}, which is only trained at a linear output layer \cite{jaeger_echo_2001-1,maass_real-time_2002} and bypassing the excessive cost of training large \acp{rnn}.
The unique properties of a quantum platform, such as the exponentially scaling phase space with respect to system size, superposition, entanglement, and the ability to process native quantum data \cite{ghosh_reconstructing_2021,gotting_exploring_2023,nielsen_quantum_2010}, make \ac{qrc} a promising architecture in the rapidly advancing field of quantum machine learning (QML).
Like their classical counterparts, quantum reservoir computers acting on bi-infinite time series data \cite{sugiura_nonessentiality_2024} rely on the so-called \textit{fading memory property} \cite{carroll_optimizing_2022} to not depend on inputs arbitrarily far in the past.
Initial concepts of \ac{qrc} models utilized a projective erase-and-write map for input encoding to introduce dissipation to otherwise ideal quantum systems as a means to fade out past information \cite{fujii_harnessing_2017,martinez-pena_information_2020,gotting_exploring_2023,palacios_role_2024}.
While this approach provides a theoretically feasible scheme for \ac{qrc}, it suffers from performance losses in the presence of additional dissipation due to imperfections in more realistic, \ac{nisq} reservoir computers based on open quantum systems.
To address this problem, newer models -- like the one proposed in this Letter -- use unitary input encodings in combination with a quantum system with tunable dissipation, resulting not only in an operationally more robust setup, but also in a larger parameter space to optimize the model's performance \cite{monzani_leveraging_2024,olivera-atencio_benefits_2023,sannia_dissipation_2024,domingo_taking_2023}.

Until now, the benchmarking and analysis of QRC performance have relied predominantly on information-theoretic metrics, such as the \ac{stmc} \cite{han_revisiting_2021,sannia_skin_2024,cindrak_enhancing_2024}, which do not take into account physical properties of the reservoir itself.
This separation has precluded a systematic understanding of how concrete physical properties, such as dissipation and absorption, govern the fundamental capabilities and practical optimization of quantum machine learning systems.

\begin{figure}
  \includegraphics{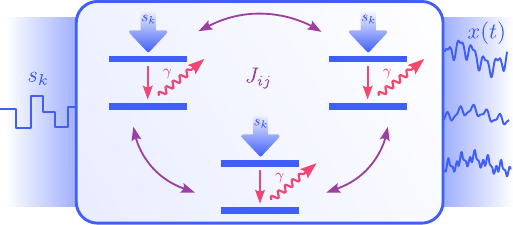}
  \caption{Illustration of the QRC paradigm using a three-qubit reservoir with the coupling matrix elements $J_{ij}$, individual qubit decay $\gamma$, and a coherent input-injection pump encoding the signal strength $s_k$. Throughout this work, qubit decay and the pump strength are equal for all qubits. For the 3-qubit case depicted here, we obtain a three-dimensional output function $x(t)$.}
  \label{fig:scheme}
\end{figure}

In this Letter, we uncover a quantitative link between a tangible, measurable physical property -- the optical absorption of a dissipative quantum many-body system -- and its resulting memory capacity for time series tasks.
Focusing on a fully-connected \ac{tfim} under coherent drive and dissipation, we show that the regime of maximal absorption coincides with maximal memory capacity, directly tying physical response to information processing ability.
This correspondence provides an intuitive explanation for the widely observed “sweet spot” of QRC performance as a function of dissipation \cite{gotting_exploring_2023,monzani_leveraging_2024,kurokawa_quantum_2024}, uniting separate domains of QML benchmarking and quantum optics.
Our results enable experimentalists to optimize QRC by tuning physical parameters accessible in the laboratory, and open the door for new physics-informed benchmarks for quantum hardware in the context of machine learning.

\section{Dissipative Quantum Systems for QRC}
\begin{figure*}
  \includegraphics{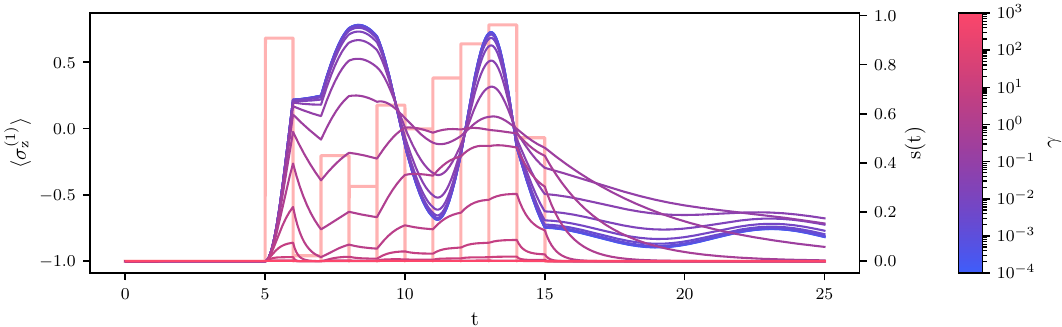}
  \caption{Time evolution of the $\sz$ expectation value of the first qubit in a three-qubit TFIM for different qubit decay strengths $\gamma$. In light red (right axis) the input signal is displayed.}
  \label{fig:time_evo}
\end{figure*}

To investigate the effects of dissipation in \ac{qrc}, we use the fully-connected \ac{tfim} \cite{pfeuty_ising_1971-1}
\begin{equation}
  H = \sum_{i<j} J_{ij} \sx^{(i)}\sx^{(j)} + h\sum_i \sz^{(i)}
  \label{eq:tfim}
\end{equation}
as the physical platform for computation, because it incorporates a sufficient amount of complexity to simulate non-trivial qubit networks while still being computationally accessible.
The qubit energy $h$ is typically set to 1, while the coupling matrix $J_{ij}$ is chosen randomly and then scaled to a certain spectral radius $J_0$, which we refer to as the \textit{coupling strength}.
In order to allow for sufficient quantum dynamics during each input cycle, the coupling strength has to be at least of the same order of magnitude as the input cycle length.
Here, the input cycle length is 1, and we set the coupling strength  to $J_0 = 0.5$ for all following calculations.
The expectation values of the $\sz$ observable of each qubit form the reservoir computer's $N$-dimensional readout function $x(t)$, where $N$ is the number of qubits in the system.
We sample the readout function at each timestep $k$ and subsample $V$-times between inputs to implement a time multiplexing.
In addition to these signals, a bias term of 1 is added to the readout, resulting in the full \textit{state collect matrix} $\mathbf{X} \in \mathbb{R}^{k \times NV + 1}$ that includes all recorded expectation values and the bias.
The output vector $\mathbf{y}$ is then constructed from the state collect matrix by applying a linear weight vector $\mathbf{w}$ that minimizes the $L^2$ distance to the target vector $\mathbf{\hat{y}}$, i.e, $\mathbf{w} = \argmin_{\mathbf{w}} |\mathbf{\hat{y}} - \mathbf{w}\mathbf{X}|^2$.

In many previous approaches \cite{fujii_harnessing_2017,martinez-pena_information_2020,gotting_exploring_2023}, the computational input is injected into the system via an erase-and-write map $\rho \mapsto \rho_\text{in}(s_k) \otimes \Tr_1 \rho$, which resets a designated input qubit to a new state parametrized by the input value $s_k$.
The drawback of this type of input encoding is that it introduces a fixed amount of dissipation to the process, which acts on top of the natural dissipation of the \ac{nisq} hardware, hindering the possibility for dissipation engineering.
Here, we choose as the input medium a coherent light source that couples resonantly to the qubits in the \ac{tfim} and is amplitude modulated according to the input signal, therefore causing no inherent dissipation.
This setup allows for complete control over the dissipation strength.
To avoid the computationally costly explicit modeling of the light field, an effective $\sy$-term is used to emulate the dynamics due to the coherent input encoding.
In the rotating frame of the pump, which we use to eliminate the impact of the qubit energy $\sz$ at resonant excitation, we obtain the time-dependent Hamiltonian
\begin{equation}
  H(t) = \sum_{i<j} J_{ij} \left(\sp^{(i)} \sm^{(j)} + \sm^{(i)} \sp^{(j)}\right) + s(t) \sum_i \sy^{(i)}.
  \label{eq:tfim_td}
\end{equation}
with $s(t) = s_{\lfloor t\rfloor}$.

As this setup so far features only unitary dynamics, it does not exhibit  fading memory and cannot be used as a reservoir computer.
Real \ac{nisq} systems are subject to various types of dissipation that force the time evolution into a steady state, which -- when unique -- is independent of the initial state, making such systems suitable for \ac{rc}.

To make the analysis more concise, we restrict ourselves to one specific dissipation mechanism, namely the qubit decay.
We model its impact on the \ac{tfim} time evolution with the \ac{gksl} equation \cite{manzano_short_2020}, consisting of the unitary von-Neumann dynamics and a dissipative part that here describes the qubit decay.
The full dynamics of the density operator is then given by
\begin{equation}
  \dot{\rho} = -\ii [H(t), \rho] + \sum_i \gamma_i \left(\sm^{(i)}\rho \sp^{(i)} - \frac{1}{2}\left\{\sp^{(i)}\sm^{(i)}, \rho\right\}\right),
  \label{eq:lindblad}
\end{equation}
where $\gamma_i \equiv \gamma$ is the decay rate of the qubits, and $\sm^{(i)}$ ($\sp^{(i)}$) is the lowering (raising) operator of qubit $i$.
Because the reservoir dissipation occurs naturally and not solely when a new input is injected, it is not required to introduce a washout sequence into the process for common-signal-induced synchronization \cite{inubushi_characteristics_2021}.
Instead, the dissipative quantum reservoir computer assumes a known steady state to arbitrary precision after a certain amount of waiting time.
To calculate this steady state, we first reformulate Eq.~\eqref{eq:lindblad} to Fock-Liouville space, yielding a Liouvillian superoperator $\mathcal{L}$, such that
\begin{equation}
  \dot{\rho} = \mathcal{L}\rho.
  \label{eq:lindblad_fls}
\end{equation}
In this new representation, the complex mathematical structure of Eq.~\eqref{eq:lindblad} turns into a linear, ordinary differential equation, which can be solved for the steady-state condition $\mathcal{L}\rho\ind{ss} = 0$.

We showcase the time evolution of a 3-qubit dissipative quantum reservoir computer in Fig.~\ref{fig:time_evo}, where the $\sz$ expectation value of the first qubit is displayed for varying qubit decays and a coupling strength of $0.5$.
To test the steady-state behavior, we leave the excitation Hamiltonian off for the first five time steps. Indeed, a constant time evolution indicative of the correct steady-state solution is found.
We follow the zero-input stage by ten randomly chosen inputs and then turn the signal off again for ten time steps, in which we observe the ``free'' time evolution of the \ac{tfim}.

Upon injection of the first nonzero input signal, the system is driven out of its steady state to a varying extent, depending on the qubit decay.
For very large decays, the deviation from the zero-input steady state is negligible, especially when shot noise is added to the simulation.
In cases where the decay itself becomes negligible compared to the timescales of the model, the unitary dynamics dominate and we observe strong oscillations in the output signal, making the system again unusable for \ac{qrc}.
Already from this analysis, one can infer that the regime interesting for \ac{qrc} should be in the range $10^{-1} < \gamma < 10^1$, where the system gets close to a new steady state that is \emph{distinct} from the zero-input steady state during each input injection period, but does not quite reach it.
In the next section, we analyse more closely how dissipation affects the system performance.

\section{Bounds of Dissipation}
A system's key property for \ac{rc} is the ability to (non-)linearly process past inputs, also called \textit{\acf{stmc}} \cite{jaeger_short_2001-1}.
In its linear form, the corresponding task consists of recovering past inputs $s_{k-\tau}$ from the current output states $\mathbf{X}_k$.
This concept has already been enhanced and extended to more sophisticated measures, including the \ac{ipc} \cite{dambre_information_2012} or the maximally well reproducible functions of the reservoir -- the so-called \textit{eigentasks} \cite{hu_tackling_2023,polloreno_limits_2023}.

Here, it suffices to consider a reduced form of the \acp{ipc}, which quantifies the ability of the QRC to reproduce legendre polynomials of the input sequence up to a certain degree.
The quality of the reproduction for each degree $i$ and delay $\tau$ is measured by the squared Pearson correlation coefficient
\begin{equation}
  C^{\tau}_{i} = \frac{\cov^2(\mathbf{y}, \mathbf{\hat{y}}_{i})}{\sigma^2(\mathbf{y})\sigma^2(\mathbf{\hat{y}}_{i})},
  \label{eq:stmc}
\end{equation}
where $\sigma(\cdot)$ is the standard deviation and $\cov(\cdot)$ the covariance. We denote the target sequence $(P_i(s_{k-\tau}), P_i(s_{k-\tau-1}), P_i(s_{k-\tau-2}), \dots )$ as $\mathbf{\hat{y}}_i$, where $P_i(\cdot)$ is the $i$-th Legendre polynomial rescaled to the interval $[0, 1]$.
The length of the training and testing sequences are both fixed to 1000 steps.
As our \ac{rc} system exhibits a dynamically fading memory, the capcities $C_i^\tau$ need to vanish for large $\tau$, allowing us to sum up the STMCs for all delays up to a certain $\tau\ind{max}$ to define a \textit{total memory capacity} $C_i := \sum_{\tau = 0}^{\tau\ind{max}} C_i^\tau$ for each degree $i$.

Finding sensible values for $\tau\ind{max}$ is not a trivial task, as the correlation between two random functions is nonzero in general, resulting in a noise tail of the capacities $C^\tau_i$ even for very large $\tau$.
To quantify this noise floor, we correlate the reservoir output $\mathbf{y}$ with a random target function, such that the thereby obtained capacities do not represent an actual memory effect, but rather random correlations.
We note that these correlations depend only on the trained weights and network architecture, not the delay $\tau$.
The process of calculating these ``noise capacities'' is repeated 500 times and we take the largest resulting capacity to define the \textit{noise threshold} $C_i^{\,\text{thresh}}$ for each degree $i$.
Once a value in the sequence $(C_i^\tau)$ drops below this noise threshold, we cut the sequence at that $\tau$ and receive the maximum delay $\tau\ind{max}$.

\begin{figure}
  \includegraphics{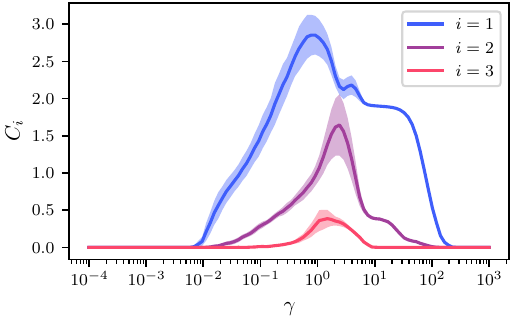}
  \caption{First three degrees of the \ac{stmc} of 15 randomly sampled 3-qubit systems for different qubit decays.}\label{fig:stmc}
\end{figure}

With the framework set, results for the total memory capacities of degrees 1, 2, and 3 for varying strengths of the qubit decay are shown in Fig.~\ref{fig:stmc}.
For small decays, where the system evolves approximately unitarily, the QRC is not able to recall any of the past inputs, so that $C_i^\tau \equiv 0$.
Interestingly, the linear \ac{stmc} displays a strong increase at $\gamma \gtrsim 10^{-2}$ with the higher-order degrees following shortly after.
Indeed, as proposed in the last section, we then find a regime of optimal memory capacity at intermediate level of dissipation before, at high qubit decay rates $\gamma \gtrsim 10^2$, the capacities drop to zero.
As higher degrees follow a similar trend, but exhibit lower capacities overall in this setup, we restrain the following discussion to the linear \ac{stmc}.
We also note that the data provided was obtained by simulating shot noise corresponding to $10^6$ shots.
For the infinite-shot ideal case, results are discussed in the Supplementary Material.

\section{Connection to physical reservoir properties}
While we gave a general intuition for the behavior of the \ac{stmc} in Fig.~\ref{fig:time_evo}, there is typically not much insight obtained about the underlying processes purely from numerical simulations of the reservoir dynamics.
By virtue of the reservoir computer being an open quantum system with a modulated light field as the input injection mechanism, we can, however, facilitate an understanding of the sweet-spot behavior via the analysis of actual physical properties of the qubit network.
Here, we consider the optical absorption $\alpha$ of the reservoir as the main property driving its capability to remember.
According to linear response theory, the frequency-dependent absorption of a network of two-level systems is given by the normalized dipole correlation's Fourier spectrum \cite{chang_non-markovian_1993,mcquarrie_statistical_2000}
\begin{align}
  \alpha_{s, \gamma}(\omega) &= \Re\int_0^\infty \e^{-\ii\omega t} \frac{\Tr[\Sx \Sx(t) \rho\ind{ss} ]}{\Tr [\Sx^2 \rho\ind{ss}]} \d t \nonumber \\
  &= \Re\int_0^\infty \e^{-\ii\omega t} \frac{\Tr[\Sx \e^{\mathcal{L}[s, \gamma]t} \Sx \rho\ind{ss} ]}{\Tr [\Sx^2 \rho\ind{ss}]} \d t,
  \label{eq:alpha}
\end{align}
where $\Sx = \sum_i \sx^{(i)}$ is the total dipole moment operator.
Here, the additional dependence of the absorption on the signal strength $s$ is explicitly marked, as the external field strongly influences the Hamiltonian structure and thus the absorption spectrum.
This circumstance requires the calculation of absorption spectra over the whole range of signal strengths, which are then averaged to get a mean response of the \ac{tfim} to optical excitation.
We justify this by the fact that the 1000 testing input signals are uniformly chosen, such that no specific signal strength has a prominent effect on the resulting \acp{stmc}, but rather the average ensemble of signals.

In order for the integral in Eq.~\eqref{eq:alpha} to converge, the trace term needs to vanish for large $t$, which is, however, prevented by the external pumping that results in a constant tail of the correlation function.
This effect was attributed by Mollow to elastic scattering of the pump light in \cite{mollow_power_1969} and leads to a delta peak in the absorption spectrum at the pump frequency.
As the delta peak is only an artifact of the monochromatic light source, we remove it by subtracting the value of the constant tail of the correlation function to obtain a meaningful absorption spectrum.
This is illustrated in Fig.~\ref{fig:SM} in the Supplementary Material for $s=1$ and different qubit decays.

\begin{figure}
  \includegraphics{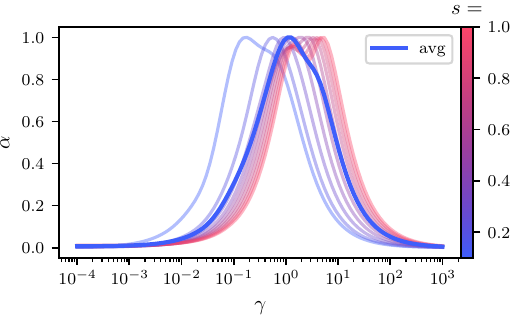}
  \caption{Resonant absorption $\alpha_{s, \gamma}(0)$ of the same 15 randomly sampled 3-qubit system as in Fig.~\ref{fig:stmc} for different signal strengths $s$ and qubit decays $\gamma$. The thick blue line marks the average over all light curves.}
  \label{fig:absorption}
\end{figure}
We investigate the resonant absorption $\alpha_{s, \gamma}(0)$, which corresponds to the qubit system's absorption at the pump frequency, in Fig.~\ref{fig:absorption}. 
It is shown for 10 different values of $s$ and varying qubit decay (color coded).
It becomes evident that the signal strength -- which was discussed before to have a considerable influence on the absorption spectrum -- in fact shifts the absorption peak by several orders of magnitude in the qubit decay.
This effect appears to be most prominent for smaller signal strengths up to about 0.5, after which the $\sy$-term in the Hamiltonian becomes dominant, and increasing it further does not change the dynamics structurally (note that a coupling strength of 0.5 was used for these results).
Notably, the absorption exhibits shoulders for some signal strengths, which is due to asymmetric shifts of the absorption spectrum's peaks, as further explained in the Supplementary Material.
The input-dependent behavior of the absorption spectrum hints at the possibility to tune either the qubit decays individually, the qubit couplings, or the signal-strength range to maximize the absorption.
We average over all signal strengths in the above-described way to receive an average absorption $\bar{\alpha}_\gamma$ representing the mean response of the \ac{tfim} to excitation.
In Fig.~\ref{fig:absorption}, this average response is shown as a thick blue line.
Remarkably, it exhibits a qualitatively very similar behavior to the \ac{stmc} as a function of dissipation in Fig.~\ref{fig:stmc}: We find a rising edge starting from around $\gamma = 10^{-2}$, developing to a maximum at a decay of approximately 1 before decreasing again to vanish completely at $\gamma \approx 10^2$.
The striking similarities between \ac{stmc} and optical absorption at the extremes of very strong and very weak amounts of dissipation follow the intuition that the system cannot possibly remember any information if it is not able to absorb it in the first place.
From the data between the extremes, and especially in the sweet spot regime around the peak, we deduce that there exists a significant correlation between the two quantities.
The observed behavior aligns with our physical understanding that the reservoir computer is more effectively supplied with information in the strong-absorption regime and should, consequently, also be able to better reproduce it.
We have verified that this also holds for systems with different coupling strengths and topologies, and results for a 4-qubit reservoir computer are shown in the Supplementary Material.
These results reveal a link between two -- at first glance -- completely unrelated properties, the first one being the information theoretical \ac{stmc}, and the second one the physical property of the optical absorption.

On a deeper level, this link is reflected in the mathematical structure of these two quantities, which both are calculated in terms of two-time correlation functions of an initial input to later system states, the Pearson correlation here being proportional to $\langle s_k \mathbf{y}(t) \rangle_k$.
A direct formal connection between the autocorrelation in Eq.~\eqref{eq:alpha} to the Pearson correlation defining the \ac{stmc} is, however, not straightforward in this framework, mainly because the training process adds an element of external control that makes it difficult to establish a mathematical connection to the  absorption, which is an intrinsic property, and further work will be necessary in this direction.

\section{Conclusion}
Our work reveals a previously unidentified link between physics and computation benchmarks in quantum reservoir computers: memory performance of coherently driven quantum reservoir computers is governed by the system's susceptibility to the input signal, here given by the optical absorption.
In demonstrating how optimal dissipation maximizes both absorption and short-term memory performance simultaneously, we provide fundamental insight into the role of dissipation in QRC and elucidate a way, in  which it can be used as an experimentally accessible tuning knob for engineering high-performing and task-specific quantum neural networks.
In this sense, we argue that the understanding of physical properties, not just abstract information measures, is important to guide quantum technology development.
There already exists a multitude of proposed or implemented physical systems for \ac{qrc}, such as photonic systems \cite{garcia-beni_scalable_2023}, nonlinear oscillators \cite{govia_quantum_2021-1}, and arrays of Rydberg atoms \cite{bravo_quantum_2022-1}, where we expect similar relations as the one between absorption and memory presented in this work.
Arrays of Rydberg atoms in particular are a promising physical system for our work, as they directly implement the \ac{tfim} on physical hardware and have proven remarkable scaling in recent experiments \cite{ebadi_quantum_2021}.
We anticipate that our results will motivate new experimental studies of tunable QRC platforms and foster the development of performance metrics grounded both in physics and information theory.

\paragraph{Acknowledgements}
We gratefully acknowledge funding by the Deutsche Forschungsgemeinschaft (German Science Foundation, DFG) via the project PhotonicQRC (Gi1121/6-1).

\bibliography{qrc_dissipation_paper.bib}

\clearpage
\onecolumngrid

\appendix
\section{SUPPLEMENTARY MATERIAL \vspace{.5cm}}

\section{Absorption spectra}
\begin{figure}[H]
  \includegraphics{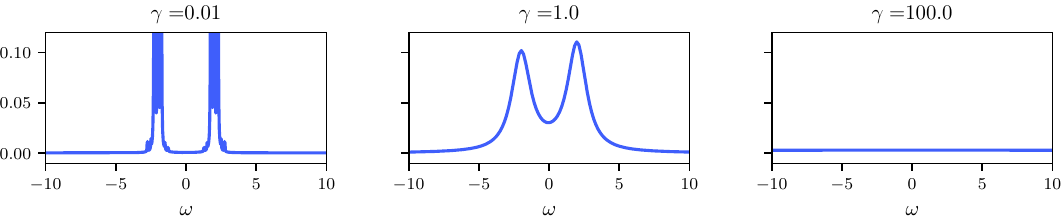}
  \caption{Absorption spectra for qubit decays $\gamma \in \{10^{-2}, 10^0, 10^2\}$ and an input signal strength of $s=1$. The zero-frequency components correspond to points on the absorption curve for $s=1$ in Fig.~\ref{fig:absorption}}
  \label{fig:SM}
\end{figure}

To gain a better insight regarding the structure of the absorption curves in Fig.~\ref{fig:absorption} in the main text, we here show three exemplary absorption spectra for the input strength $s=1$ and three different qubit decays in Fig.~\ref{fig:SM}.
For a small decay ($\gamma = 10^{-2}$), we find sharp peaks centered around the zero-frequency in the absorption spectrum, which are a result of the external pumping and change in position for different input signals.
Upon increasing the dissipation strength, the typical line broadening occurs and allows for the absorption of information encoded in the pump frequency of $\omega = 0$ -- it is around this dissipation that the absorption, and with it the \ac{stmc}, is maximized.
Elevating the dissipation even further to $\gamma = 100$ leads to an almost completely flat spectrum, as the line broadening washes away all structure in the spectrum.

\begin{figure}[H]
  \includegraphics{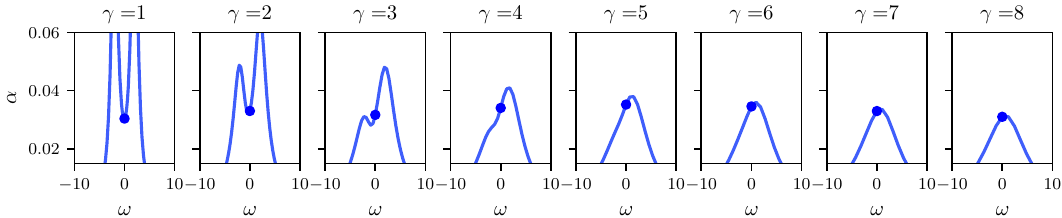}
  \caption{Absorption spectra for a 3-qubit reservoir computer with a coupling strength $J_0 = 0.5$ exposed to a signal of amplitude $s = 1$ for a range of qubit decays. The $\alpha$-axis is cropped to make the changes of the resonant absorption (blue dots) more visible. We observe a non-monotonous behavior of the absorption with respect to the decay rate.}\label{fig:sm_abs}
\end{figure}

Furthermore, the absorption curves in Fig.~\ref{fig:absorption} exhibit shoulders around the maximum absorption.
To gain insight into this observation, in Fig.~\ref{fig:sm_abs}, we take as an example a single 3-qubit reservoir with a coupling strength of $J_0 = 0.5$ exposed to a signal of amplitude $s = 1$ for a range of qubit decays.
For the smallest decay (here, $\gamma = 1$), the resonant absorption (i.e. absorption at $\omega = 0$ in the rotating frame) of the \ac{tfim} lies in between two peaks.
As the peaks merge asymmetrically for increasing decays, the resonant absorption behaves non-monotonically until the peaks are fully merged at around $\gamma = 5$, after which the absorption decreases steadily.
This behavior gives rise to the shoulders appearing in Fig.~\ref{fig:absorption} and strongly depends on the specific coupling and signal strength.
It cannot trivially be bypassed by changing the drive frequency, as this also changes the peak positions.

\section{Infinite shot STMC}
\begin{figure}[H]
  \centering \includegraphics{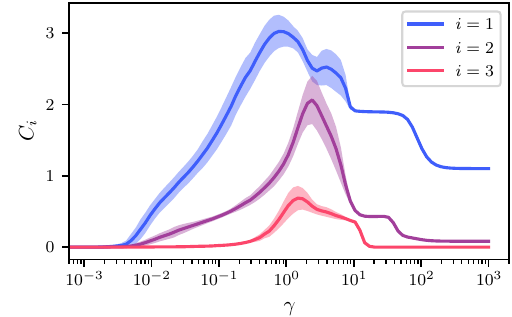}
  \caption{The first three \acp{stmc} for varying qubit decay strength and zero shot noise. As opposed to the results with finite shot noise in Fig.~\ref{fig:stmc}, the capacities do not vanish for large qubits decays.}
  \label{fig:SMinf}
\end{figure}

As discussed in the main text, Fig.~\ref{fig:stmc} depicts the \ac{stmc} for finite shot noise corresponding to $10^6$ measurements.
In the infinite-shot case, which is shown here in Fig.~\ref{fig:SMinf}, we observe a slightly different behavior of the \ac{stmc} for larger decays: Even the most minute deviations from the zero-input steady state can be utilized by an ideal, completely noise-free reservoir computer to extract the current input, resulting in a linear \ac{stmc} of almost exactly one.
We find that higher orders of the \ac{stmc} do not benefit to the same degree from the absence of noise and still drop close to zero for large decays.

\section{STMC to absorption relation generality}
\begin{figure}[H]
  \includegraphics[width=\textwidth]{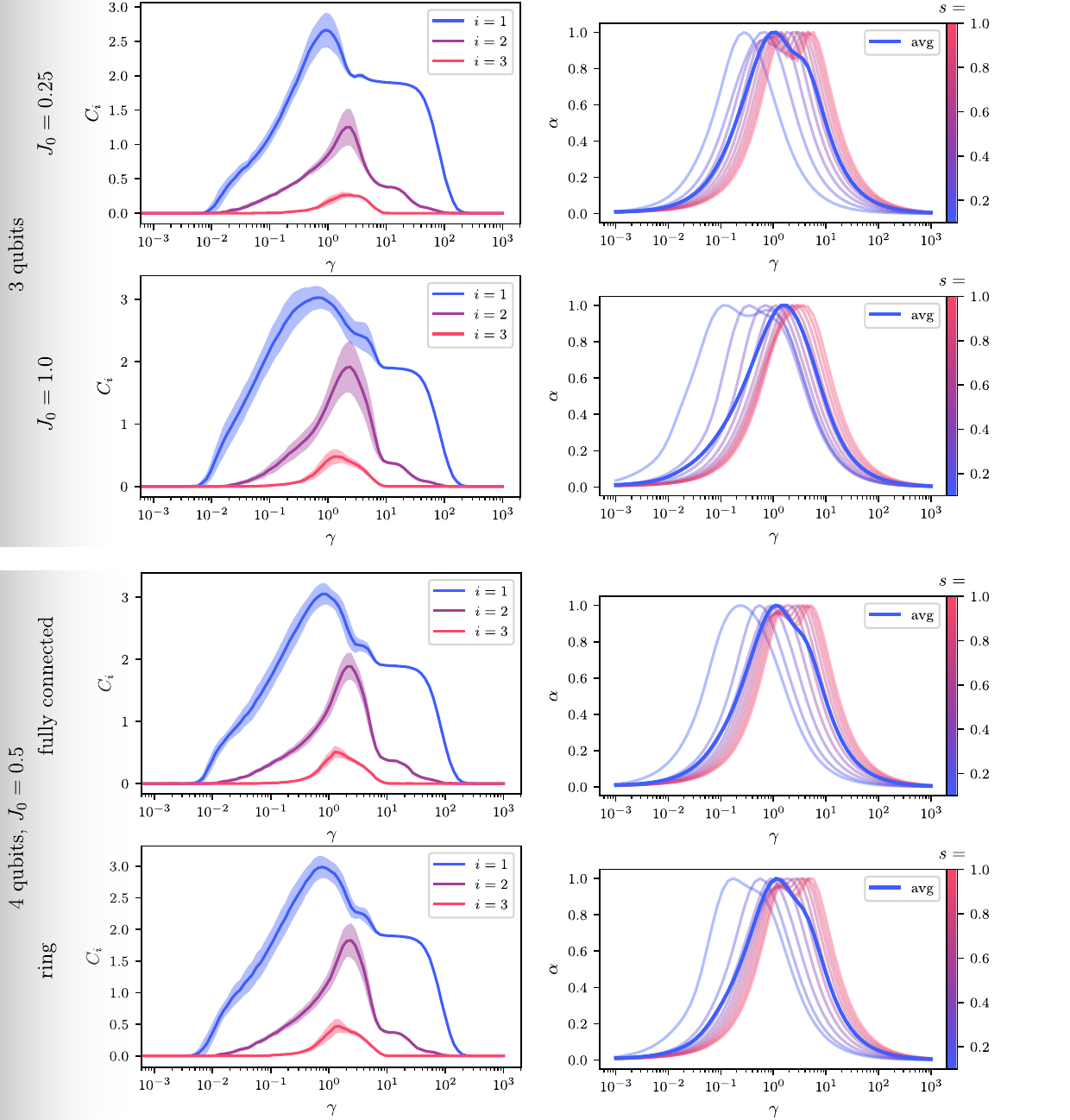}
  \caption{The \ac{stmc} and absorption for different coupling strengths, system sizes, and topologies. We observe the same behavior as for the systems considered in the main text.}\label{fig:stmc_more_q}
\end{figure}

In Fig.~\ref{fig:stmc_more_q}, we show the same simulations as carried out in the main text for 3 qubits in Figs.~\ref{fig:stmc} and \ref{fig:absorption}, but for different coupling strengths $J_0 \in \{0.25, 1.0\}$ and two 4-qubit reservoirs with all-to-all and ring topology, respectively.
The results remain qualitatively the same as for the systems considered in the main text, indicating the generality of the revealed connection between absorption and memory capacity also for larger systems, different topologies, and coupling strengths.

\begin{acronym}
  \acro{qrc}[QRC]{quantum reservoir computing}
  \acro{qelm}[QELM]{quantum extreme learning machine}
  \acro{rc}[RC]{reservoir computing}
  \acro{tfim}[TFIM]{transverse-field Ising model}
  \acro{nisq}[NISQ]{noisy intermediate-scale quantum}
  \acro{gksl}[GKSL]{Gorini–Kossakowski–Sudarshan–Lindblad}
  \acro{ipc}[IPC]{information processing capacity}
  \acro{stmc}[STMC]{short-term memory capacity}
  \acro{rnn}[RNN]{recurrent neural network}
\end{acronym}
\end{document}